\documentclass[a4paper,twoside]{article}
\newif\ifdraft \global\drafttrue
\def\production{\global\draftfalse}
\production
\usepackage[T1]{fontenc}
\usepackage[latin1]{inputenc}
\usepackage{a4wide}
\usepackage{times}
\usepackage{graphicx}
\usepackage{theorem}
\usepackage[colorlinks=true,urlcolor=blue]{hyperref}
\usepackage{color}
\usepackage{fancyhdr}
\usepackage{latexsym}
\usepackage{amsmath}
\usepackage{bbm}
\usepackage{amssymb}
\usepackage{amsfonts}
\usepackage{enumerate}
\usepackage{cancel}
\ifdraft
\usepackage{showkeys}
\fi
\newcounter{smallarabics}

\newcounter{smallroman}

\newcommand{\ben}{\begin{enumerate}[{\rm (1)}]}
\newcommand{\een}{\end{enumerate}}

\newtheorem{theoreme}{Theorem }[section]
\newtheorem{proposition}[theoreme]{Proposition}
\newtheorem{lemma}[theoreme]{Lemma}
\newtheorem{definition}[theoreme]{Definition}

\def\rr{{\mathbb R}}

\def\cc{{\mathbb C}}

\def\textsl{{}}

\def\Re{{\rm Re}\,}
\def\ch{\mathfrak{h}}

\def\sa{{\rm sa}}

\def\c0inf{C_0^\infty}
\def\bep{\begin{proposition}}
\def\eep{\end{proposition}}

\def\cH{{\cal  H}}

\def\cR{{\cal R}}

\def\cB{{\cal B}}

\def\f{{\rm f}}

\def\i{{\rm i}}
\newcommand{\beq}{\begin{equation}}
\newcommand{\eeq}{\end{equation}}
\newcommand{\bear}[1]{\begin{array}{#1}}
\newcommand{\ear}{\end{array}}

\newcommand{\e}{\mathrm{e}}
\renewcommand{\i}{\mathrm{i}}

\renewcommand{\d}{\mathrm{d}}

\def\qed{$\Box$\medskip}
\def\cP{{\cal P}}

\def\cN{{\cal N}}

\def\cW{{\cal W}}

\def\cQ{{\cal Q}}
\def\cK{{\cal K}}

\def\bel{\begin{lemma}}
\def\eel{\end{lemma}}
\def\bet{\begin{theoreme}}
\def\eet{\end{theoreme}}
\def\bed{\begin{definition}}
\def\eed{\end{definition}}
\def\bar{\overline}

\def\12{\frac{1}{2}}

\def\e{{\rm e}}

\def\d{{\rm d}}
\def\Ran{{\rm Ran}\,}

\def\one{{\mathbbm 1}}

\def\cH{{\cal H}}

\def\Ker{{\rm Ker}\,}
\def\Dom{{\rm Dom}\,}

\def\cS{{\cal S}}
\def\cR{{\cal R}}

\def\A{{\rm A}}

\def\fM{\mathfrak{M}}
\def\fH{\mathfrak{H}}

\def\tr{{\rm tr}}

\def\cO{\mathcal{O}}

\def\Jl0{J^{(l)}_{0}}
\def\Jr0{J^{(r)}_0}

\def\dl0{\delta_{-1}}
\def\dr0{\delta_{1}}

\def\chl0{\chi^{(l)}_0}
\def\chr0{\chi^{(r)}_0}
\def\ch0lr{\chi^{(l/r)}_0}
\def\de0{\delta_0}

\def\Jlr0{J^{(l/r)}_0}

\def\ch{{\rm ch}}

\begin{document}
\def\today{}
\title{A note on the Landauer principle\\ in quantum statistical mechanics}
\author{Vojkan  Jak\v{s}i\'c$^{1}$ and Claude-Alain  Pillet$^{2}$
\\ \\ 
$^1$Department of Mathematics and Statistics, 
McGill University, \\
805 Sherbrooke Street West, 
Montreal,  QC,  H3A 2K6, Canada
\\ \\
$^2$Universit\'e de Toulon, CNRS, CPT, UMR 7332, 83957 La Garde, France\\
Aix-Marseille Universit\'e, CNRS, CPT, UMR 7332, 13288 Marseille, France\\
FRUMAM
}
\maketitle
{\small
{\bf Abstract.} The Landauer principle asserts that the energy cost of erasure
of one bit of information by the action of a thermal reservoir in equilibrium at
temperature $T$ is never less than $k_{\rm B}T\log 2$. We discuss Landauer's 
principle for quantum statistical models describing a finite level quantum 
system $\cS$ coupled to an infinitely extended thermal reservoir $\cR$.
Using Araki's perturbation theory of KMS states and the Avron-Elgart
adiabatic theorem we prove, under a natural ergodicity assumption
on the joint system $\cS+\cR$, that Landauer's bound saturates for adiabatically 
switched interactions. The recent work~\cite{ReWo} on the subject is discussed and compared.} 
\thispagestyle{empty}
\section{Introduction} 
Consider a quantum system $\cS$ described by a finite dimensional Hilbert 
space $\cH_\cS$. Initially, $\cS$ is in a state described by a density 
matrix $\rho_\i$. Let $\rho_\f$ be another density matrix on $\cH_\cS$.
The Landauer principle~\cite{La,Ma} sets a lower bound on the energetic cost 
of the state transformation $\rho_\i\to\rho_\f$ induced by the action of a
reservoir $\cR$ in thermal equilibrium at temperature $T$. The principle can 
be derived from the second law of thermodynamics, provided one accepts that
the (Clausius) entropy of the system $\cS$ in the state $\rho_{\i/\f}$
coincides with its von~Neumann entropy
\[
 S(\rho_{\i/\f})=-k_{\rm B}\tr(\rho_{\i/\f}\log\rho_{\i/\f}).
\]
Since this is only correct if both $\rho_{\i/\f}$ are equilibrium states,
such a derivation puts severe limits on the domain of validity of the
Landauer principle, in contrast to its supposed universality
and experimental verifications~\cite{Ber}.

The derivation goes as follows. The {\sl decrease} in the entropy of the system $\cS$ in the transition
$\rho_\i\to\rho_\f$ is  
\[
\Delta S=S(\rho_\i)-S(\rho_\f).
\]
Let $\Delta\cQ$ denote the {\sl increase} in the energy of the reservoir $\cR$ in 
the same process. Assuming that the joint system $\cS+\cR$ is isolated and that 
the reservoir $\cR$ is large enough to remain in equilibrium at temperature $T$ 
during the whole process, the entropy of $\cR$ increases by 
$\Delta S_\cR=\Delta\cQ/T$ and the entropy balance equation of the process
(see~\cite{dGM}) reads
$$
\Delta S+\sigma=\frac{\Delta\cQ}{T},
$$
where $\sigma$ is the entropy produced by the process. The second law of 
thermodynamics stipulates that $\sigma\geq0$, with equality iff the 
transition is the result of a reversible quasi-static process. Hence, the inequality
$$
\Delta\cQ\geq T\Delta S
$$
holds for arbitrary processes, with equality being achieved by reversible 
quasi-static processes in which the change of the total entropy vanishes
$$
\Delta S_{\rm tot}=-\Delta S+\Delta S_\cR=0.
$$

With $d=\dim\cH_\cS$, $S(\rho_\i)$ is 
maximal and equal to $k_{\rm B}\log d$ if $\rho_\i=\one/d$ is the chaotic state
and $S(\rho_\f)$ is minimal and equal to $0$ if 
$\rho_\f=|\psi\rangle\langle \psi|$ is a pure state. It follows that 
\beq\label{lan-pri}
\Delta\cQ\geq k_{\rm B}T\log d.
\eeq
If in addition  $d=2$, then $\Delta\cQ$ is the energy
cost of the erasure of the qubit of information stored in $\rho_\i$ 
and~\eqref{lan-pri} reduces to the Landauer bound. 

The defects of the above ``derivation'' of the Landauer principle are manifest.
In spite of its importance, there are very few mathematically rigorous results  
concerning the derivation of the Landauer bound from the first principles of 
statistical mechanics. 

In an interesting recent work, Reeb and Wolf~\cite{ReWo} point out that the lack 
of mathematically precise formulation and proof of the Landauer principle in 
the context of quantum statistical mechanics has led to a number of 
controversies in the literature regarding its nature and validity. To remedy this 
fact, in the same work they provide a derivation of the Landauer principle which 
we will discuss in the next section.

One  of the values of the paper~\cite{ReWo} is that it has brought the Landauer 
principle to the attention of researchers in quantum statistical mechanics.

In this note we shall examine the Landauer principle in the context of recent 
developments in the mathematical theory of open quantum systems\footnote{
We shall discuss the Landauer principle only for microscopic Hamiltonian models 
describing coupled system $\cS+\cR$. Repeated interaction systems 
(see~\cite{BJM}) are an instructive and physically important class of models in 
quantum statistical mechanics that also allow for  mathematically rigorous 
analysis of the Landauer principle. This analysis is presented in~\cite{Raq}.} 
(\cite{AS},\cite{ASF1}--\cite{ASF4}, \cite{AJPP1,AJPP2,BFS,DJ,DJP,dR,dRK,FM,FMU,FMSU,JOP1,JOP2,JOPP}, \cite{JP1}--\cite{JP7}, 
\cite{MMS1,MMS2,Pi1,Pi2,Ru1,Ru2,TM})\footnote{This is by no means a comprehensive
list of references. Some of the earlier works that motivated these developments 
are~\cite{AM,BM,LeSp,McL,Rob,PW1,Sp1,Zu1,Zu2}.}, and compare the outcome with the 
results of~\cite{ReWo}.

The paper is organized as follows. In Section~\ref{sec-RW} we will review the 
work~\cite{ReWo}. The entropy balance equation in quantum statistical mechanics 
and its implication regarding Landauer's principle are presented in
Section~\ref{sec-ep-bal}. We discuss the Landauer principle for instantaneously 
switched interactions in Section~\ref{sec-isi} and for adiabatically 
switched interactions in Section~\ref{sec-asi}. Section~\ref{sec-disc} is devoted 
to the  discussion of the results presented in this note. The proofs are given in 
Section~\ref{sec-proof}. 

This note is similar in spirit to the recent work~\cite{JOPS}. It is an attempt 
to bring together two directions of research which seem largely unaware of each 
other, in the hope that they both may benefit from this connection.

\bigskip
{\bf Acknowledgments.} The research of V.J. was partly supported by NSERC.
C.A.P. is grateful to the Department of Mathematics and Statistics at McGill 
University for its warm hospitality.

\section{The Reeb-Wolf derivation}
\label{sec-RW}

Suppose that $\cR$ is described by a finite dimensional Hilbert space $\cH_\cR$ 
(we shall call such reservoirs confined) and Hamiltonian $H_\cR$. Initially, 
$\cR$ is in thermal equilibrium at temperature $T$, and its state is described by the
density matrix  
\beq\label{Gibbs}
\nu_\i=\e^{-\beta H_\cR}/Z,
\eeq
where $Z=\tr(\e^{-\beta H_\cR})$ and $\beta=1/T$ (in the following, we shall
set Boltzmann's constant $k_{\rm B}$ to $1$). The Hilbert space of the coupled 
system $\cS+\cR$  is 
\[
\cH=\cH_\cS\otimes\cH_\cR,
\]
and its initial state has the product structure
\[
\omega_\i=\rho_\i\otimes\nu_\i.
\]
In what follows $\tr_{\cS/\cR}$ denotes the partial trace over $\cH_{\cS/\cR}$ 
and, whenever the meaning is clear within the context, we will denote operators
of the form $A\otimes\one$, $\one\otimes A$ by $A$. The relative entropy of two 
positive linear maps $\zeta_1,\zeta_2$ is defined by  
\beq\label{convention}
S(\zeta_1|\zeta_2)=\tr(\zeta_1(\log\zeta_1-\log \zeta_2)).
\eeq
If $\tr\,\zeta_1=\tr\,\zeta_2$, then $S(\zeta_1|\zeta_2)\geq 0$ with equality 
iff $\zeta_1=\zeta_2$.

Let $U:\cH\rightarrow\cH$ be a unitary operator inducing the state 
transformation 
$$
\omega_U=U\omega_\i U^\ast.
$$
The transformed states of the subsystems $\cS$ and $\cR$ are given by
$$
\rho_{U}=\tr_\cR(\omega_U),\qquad\nu_{U}=\tr_\cS(\omega_U).
$$
In the literature, the relative entropy
\[
S(\omega_U|\rho_U \otimes \nu_U)=S(\rho_U)+S(\nu_U)-S(\omega_U)
\]
is sometimes called mutual information and the fact that it is non-negative
yields the subadditivity of entropy. The decrease in entropy of $\cS$ and the 
increase in energy of $\cR$ in the transition process $\omega_\i\to\omega_U$ 
are respectively
$$
\Delta S=S(\rho_\i)-S(\rho_U),\qquad
\Delta\cQ=\tr(\nu_U H_\cR)-\tr(\nu_\i H_\cR).
$$
The unitarity of $U$ and the product structure of $\omega_\i$ imply
\[
S(\omega_{U})=S(\omega_\i)=S(\rho_\i)+S(\nu_\i),
\]
and Eq.\;\eqref{Gibbs} yields
$$
S(\nu_\i)=\beta\tr(\nu_\i H_\cR)+\log Z.
$$
It follows that
\begin{align*}
S(\omega_U|\rho_U\otimes\nu_\i)
&=-S(\omega_U)-\tr(\omega_U(\log\rho_U+\log\nu_\i))\\[2mm]
&=-S(\rho_\i)-S(\nu_\i)-\tr(\rho_U\log\rho_U)-\tr(\nu_U\log\nu_\i)\\[2mm]
&=-S(\rho_\i)-\beta\tr(\nu_\i H_\cR)-\log Z+S(\rho_U)+\beta\tr(\nu_U H_\cR)
+\log Z,
\end{align*}
and one arrives at the entropy balance equation
\beq\label{akin}
\Delta S+\sigma=\beta \Delta\cQ,
\eeq
where the entropy production term is given by
\begin{equation}\label{sigmaForm}
\sigma=S(\omega_U|\rho_U\otimes\nu_\i)\geq 0.
\end{equation}
This leads to the Landauer bound
\beq\label{bound-again}
\beta\Delta\cQ\geq \Delta S.
\eeq
Note that \eqref{sigmaForm} implies that $\sigma=0$ iff $\omega_U=\rho_U\otimes \nu_\i$.
The last relation yelds $\nu_U=\nu_\i$ and hence $\Delta\cQ=0$. Thus
equality holds in \eqref{bound-again} iff $\Delta\cQ=\Delta S=0$. In this case, it 
further follows from the identities 
$$
\tr(\rho_\i^{\alpha})\,\tr(\nu_\i^{\alpha})
=\tr(\omega_\i^{\alpha})=\tr(\omega_U^{\alpha})
=\tr(\rho_U^{\alpha})\,\tr(\nu_\i^{\alpha}),
$$
that $\tr(\rho_\i^{\alpha})=\tr(\rho_U^{\alpha})$ holds for all $\alpha\in\cc$.
One easily concludes from this fact that $\rho_\i$ and $\rho_U$ are unitarily
equivalent.

The following additional points are discussed in \cite{ReWo}.

\bigskip
{\bf Remark 1.} Given $\rho_\i$, $\beta$, $\cH_\cR$ and $H_\cR$, there are 
many target states $\rho_\f$ for which there is no unitary $U$ such that
$\rho_U=\rho_\f$. Let 
\[
\ell=e_{\rm max}-e_{\rm min},
\]
where $e_{{\rm max}/{\rm min}}$ is the maximal/minimal eigenvalue of $H_\cR$.
Then, for any unitary $U$,
\[
\e^{-\ell\beta}\rho_\i\le\rho_U\le\e^{\ell\beta}\rho_\i.
\]
This constrains the set of possible target states $\rho_\f$. To reach a given 
$\rho_\f$, either exactly or up to a prescribed small error, one may need  to 
adjust $\cH_\cR$, $H_\cR$, and $U$. The following example illustrates one trivial 
way in which $\rho_\f$ can always be reached. 

\medskip
{\bf Example 1.} Let $\rho_\f>0$ be the target state. Set $\cH_\cR=\cH_\cS$, 
$\nu_\i=\rho_\f$, $H_\cR=-\log \rho_\f$. In this example, 
$\beta=1$. Let $U$ be the flip map, $U(\phi\otimes\psi)=\psi\otimes\phi$.
Then $\rho_U=\rho_\f$, $\nu_U=\rho_\i$, the entropy production is 
\[
\sigma=S(\rho_\i|\rho_\f),
\]
and $\Delta\cQ=\Delta S$ iff $\rho_\f=\rho_\i$.

\bigskip
{\bf Remark 2.} It turns out that Inequality~\eqref{bound-again}, as a lower
bound of $\Delta\cQ$ in terms of $\Delta S$, is not optimal.
This can be seen as follows, starting with the standard bound (see for example
Theorem~1.15 in~\cite{OP})
\[
S(\omega_U|\rho_U\otimes \nu_\i)
\geq\frac12\|\omega_U-\rho_U\otimes\nu_\i\|^2_1,
\]
where $\|X\|_1=\tr\,|X|=\sup_{A\not=0}|\tr(AX)|/\|A\|$ is the  trace norm.
With $e=(e_{\rm max}+e_{\rm min})/2$, we can estimate
\[
\|\omega_U-\rho_U\otimes\nu_\i\|_1
\geq\frac{|\tr[(H_\cR-e)(\omega_U-\rho_U\otimes\nu_\i)]|}{\|H_\cR-e\|}
=\frac{|\Delta\cQ|}{\ell/2},
\]
and so  the entropy production \eqref{sigmaForm} satisfies
\beq
\sigma\ge2\left(\frac{\Delta\cQ}{\ell}\right)^2.
\label{well}
\eeq
Combining  (\ref{akin}) and (\ref{well}) and solving the resulting quadratic inequality shows that the possible
entropy changes are restricted by the constraint
$\Delta S\le S_0=\beta^2\ell^2/8$ 
and that the corresponding energy cost satisfies the improved bound
$$
\beta\Delta\cQ\ge\left(1+\frac{1-\sqrt{1-\Delta S/S_0}}%
{1+\sqrt{1-\Delta S/S_0}}\right)\Delta S.
$$

A part of the discussion in~\cite{ReWo} is devoted to the refinement and 
optimization of the estimate~\eqref{bound-again} in the spirit of the above 
argument.

\medskip
{\bf Example 2.} On physical grounds, one expects saturation of the Landauer bound 
for quasi-static reversible processes. The following toy example 
of~\cite{ReWo} illustrates this point. Let $\rho_\f>0$ be a given target state. 
Let $\rr\ni t\mapsto\rho(t)$ be a twice continuously differentiable map with 
values in density matrices on $\cH_\cS$ such that $\rho(0)=\rho_\i$,
$\rho(1)=\rho_\f$, and $\rho(t)>0$ for $t\in]0,1]$. Given a positive
integer $N$, set $\rho_n=\rho(n/N)$, $\cH_{\cR}=\bigotimes_{n=1}^N\cH_\cS$, 
$ \nu_\i=\bigotimes_{n=1}^N\rho_n$. With $\beta=1$, it follows that $H_\cR=-\sum_{n=1}^N\log\rho_n$, $\omega_\i=\bigotimes_{n=0}^N\rho_n$.
Let $U:\cH\rightarrow\cH$ be defined by 
\[
U(\psi_0\otimes\psi_1\otimes\cdots\otimes\psi_N)
=\psi_N\otimes\psi_0\otimes\cdots\otimes\psi_{N-1}.
\]
Then $\rho_U=\rho_\f$ and 
$$
\Delta\cQ_N=\tr(\nu_U H_\cR)-\tr(\nu_\i H_\cR)
=\sum_{n=1}^N\tr [(\rho_{n}-\rho_{n-1})\log \rho_{n}].
$$
The differentiability assumption allows us to rewrite the 
r.h.s.\;of the previous identity as a Riemann sum, leading to 
\[
\lim_{N\rightarrow \infty}\Delta\cQ_N
=\int_0^1\tr(\dot\rho(t)\log\rho(t))\d t
=S(\rho_\i)-S(\rho_\f).
\]
In this example the number of steps $N$ plays the  role of an adiabatic parameter 
and the limit $N\rightarrow\infty$ leads to a quasi-static process with optimal 
Landauer bound. 

\bigskip
{\bf Remark 3.} In the Landauer erasing principle $\rho_\i=\one/d$ and 
$\rho_\f=|\psi\rangle\langle\psi|$. Pure target states are thermodynamically 
singular and cannot be reached by the action of a thermal reservoir at strictly
positive temperature. It follows from Example~2 that for any $\epsilon >0$ one 
can find $\rho_\f^\prime$, $\cH_\cR$, $\nu_\i$, and $U$ such that 
$\|\rho_\f-\rho_\f^\prime\|_1<\epsilon$, $\rho_U=\rho_\f^\prime$, 
and that the energy cost of the transformation 
$\rho_\i\to\rho_\f^\prime$ satisfies $\beta\Delta\cQ\geq\log d-\epsilon$.  
The last result can be refined by considering infinite dimensional 
$\cH_\cR$'s and allowing for Hamiltonians $H_\cR$ with formally infinite energy 
levels. An additional toy example discussed in Section~6 of \cite{ReWo} 
illustrates this point. 

\bigskip
With the exception of the toy example mentioned in Remark~3, the work 
\cite{ReWo} is exclusively  concerned with finite dimensional thermal reservoirs.
The authors discuss several additional topics including  possible extensions of 
the notion of Landauer processes. The paper contains  valuable discussions and 
clarifications concerning the physics literature on the Landauer principle. 
In the final Section~7 of the paper, the authors list a number of open 
problems/conjectures, including the following, on which we will comment 
later: 
\begin{quote} {\bf Conjecture} \cite{ReWo}. {\em Landauer's Principle
can probably be formulated within the general statistical mechanical framework of 
$C^\ast$ and $W^\ast$ 
dynamical systems~\cite{BR2,PW1,Th} and an equality version akin to~\eqref{akin} 
can possibly be proven. Note that in this framework the mutual information can be
written as a relative entropy and the heat flow as a derivation w.r.t.\;the 
dynamical semigroup.}
\end{quote}

We now turn to the discussion of the Landauer principle in the context of the existing mathematical theory of open quantum systems. 

\section{The entropy balance equation}
\label{sec-ep-bal}

We start with the following remark regarding the derivation  of the previous 
section. Let $\eta=\one\otimes\nu_\i$. Then 
\[
S(\rho_\i)-S(\rho_U)+\sigma=S(\omega_U|\eta)-S(\omega_\i|\eta),
\]
and~\eqref{akin} can be written as 
\beq\label{akin-2}
S(\omega_U|\eta)-S(\omega_\i|\eta)=\beta \Delta\cQ.
\eeq
The relation~\eqref{akin-2} is a special case of the general entropy
balance equation in quantum statistical mechanics. In the form~\eqref{akin-2} it 
goes back at least to Pusz and Woronowicz (see the Remark at the end of 
Section~2 in~\cite{PW1}) and was rediscovered in~\cite{JP3,JP7,Pi1}, see 
also~\cite{LeSp,McL,O1,O2,OHI,Sp2,Ru2,TM,Zu1,Zu2} for related works on the 
subject. To describe~\eqref{akin-2} in  full generality we assume that the 
reader is familiar with basic definitions and results of algebraic quantum 
statistical mechanics, and in particular with Araki's perturbation theory 
of KMS structure. This material is standard and 
can be found in the monographs~\cite{BR1,BR2}. A modern exposition of the 
algebraic background  can be found in~\cite{BF,DJP,Pi2}. The interested reader 
should also consult the fundamental paper~\cite{HHW}.
For definiteness we will work with $C^\ast$-dynamical systems. With only 
notational changes all our results and proofs easily extend to $W^\ast$-dynamical systems 
and we leave such generalizations to the reader.

In the algebraic framework, a quantum system is described by a $C^\ast$-dynamical 
system $(\cO,\tau)$. $\cO$ is a $C^\ast$-algebra, with a unit $\one$, and
$\tau$ is a strongly continuous one-parameter group of $\ast$-automorphisms of 
$\cO$. Elements of $\cO$ are observables, and their time evolution, in the 
Heisenberg picture, is given by $\tau$. We denote by $\cO^\sa$ the set
of self-adjoint elements of $\cO$. A state of the system is a positive 
linear functional $\omega$ on $\cO$ such that $\omega(\one)=1$. It is
$\tau$-invariant if $\omega\circ\tau^t=\omega$ for all $t\in\rr$.
A thermal equilibrium state at inverse temperature $\beta$ is a 
$(\tau,\beta)$-KMS state. Such states are $\tau$-invariant.

Given a state $\omega$, the GNS construction provides a Hilbert space 
$\cH_\omega$, a $\ast$-morphism $\pi_\omega:\cO\to\cB(\cH_\omega)$\footnote{Throughout 
the paper $\cB(\cH)$ denotes the usual $C^\ast$-algebra of all bounded operators on a 
Hilbert space $\cH$.} and a unit
vector $\Omega_\omega\in\cH_\omega$ such that $\pi_\omega(\cO)\Omega_\omega$ is 
dense in $\cH_\omega$ and $\omega(A)=\langle \Omega_\omega,\pi_\omega(A)\Omega_\omega\rangle$
for all $A\in\cO$. A state given by $\zeta(A)=\tr(\rho\pi_\omega(A))$, where
$\rho$ is a density matrix on $\cH_\omega$, is said to be $\omega$-normal.
We denote by $\cN_\omega$ the set of all $\omega$-normal states on $\cO$. The state
$\omega$ is called ergodic for $(\cO,\tau)$ if, for all states 
$\zeta\in\cN_\omega$ and all $A\in\cO$,
\[
\lim_{t\rightarrow\infty}\frac1t\int_0^t\zeta(\tau^s(A))\d s =\omega(A),
\]
and mixing if 
\[
\lim_{t\rightarrow \infty}\zeta(\tau^t(A))=\omega(A).
\]

If $\omega$ is a $\tau$-invariant state, then there is a unique
self-adjoint operator $L_\omega$ on $\cH_\omega$ such that $L_\omega\Omega_\omega=0$
and $\pi_\omega(\tau^t(A))=\e^{\i tL_\omega}\pi_\omega(A)\e^{-\i tL_\omega}$ for all 
$t\in\rr$ and $A\in\cO$. $L_\omega$ is called the $\omega$-Liouvillean of the dynamical 
system $(\cO,\tau)$. If $\omega$ is a $(\tau,\beta)$-KMS state, then $L_\omega$ is
also the standard Liouvillean (see Section \ref{sec-proof}) of $(\cO,\tau)$.
     
The reservoir $\cR$ is described by a $C^\ast$-dynamical system 
$(\cO_\cR,\tau_\cR,\nu_\i)$ in thermal equilibrium 
at inverse temperature $\beta>0$. We denote by $\delta_\cR$ the 
generator of $\tau_\cR$, $\tau_\cR^t=\e^{t\delta_\cR}$ and by $L_\cR$ its
standard Liouvillean. If the reservoir is 
confined, then $\cO_\cR=\cB(\cH_\cR)$, $\delta_\cR(\,\cdot\,)=\i[H_\cR,\,\cdot\,]$, 
and $\nu_\i=\e^{-\beta H_\cR}/\tr(\e^{-\beta H_\cR})$\footnote{If $\dim\cH<\infty$, we 
shall not distinguish between positive linear functionals on $\cB(\cH)$ and positive 
elements of $\cB(\cH)$. They are identified by $\zeta(A)=\tr(\zeta A)$.}.
The GNS Hilbert space $\cH_{\nu_\i}$ is $\cO_\cR$ equipped with the inner
product $\langle X,Y\rangle=\tr(X^\ast Y)$, the morphism $\pi_{\nu_\i}$ is defined
by $\pi_{\nu_\i}(A)X=AX$, $\Omega_{\nu_\i}=\nu_\i^{1/2}$ and $L_\cR X=[H_\cR,X]$.
However, in the remainder of this note we shall be concerned with
infinitely extended reservoirs.

The $C^\ast$-algebra of the system $\cS$, described by the finite dimensional Hilbert 
space $\cH_\cS$, is  $\cO_\cS=\cB(\cH_\cS)$. The $C^\ast$-algebra of the joint system 
$\cS+\cR$ is 
$$
\cO=\cO_\cS\otimes\cO_\cR,
$$
and its initial state is
$$
\omega_\i=\rho_\i\otimes \nu_\i,
$$
where $\rho_\i$ denotes the initial state of $\cS$.
We continue with our notational convention of omitting tensored identity, hence $\delta_\cR=\mathrm{Id}\otimes\delta_\cR$, etc.

Let $S(\zeta_1|\zeta_2)$ be the relative entropy 
of two positive linear functionals $\zeta_1, \zeta_2$ on $\cO$
\cite{Ar2, Ar3}, with the ordering convention of~\cite{BR2,Don,JP3,JP6,DJP} and 
the sign convention of~\cite{Ar2,Ar3,OP} (with these conventions the relative 
entropy of two density matrices is given by~\eqref{convention}). The basic 
properties of the relative entropy are most easily deduced from  the 
Pusz-Woronowicz-Kosaki variational formula~\cite{Ko,PW2}
$$
S(\zeta_1|\zeta_2)= \sup\int_0^\infty\left[\frac{\zeta_2(\one)}{1+t}- \zeta_2(y^\ast(t)y(t))-\frac{1}{t}\zeta_1(x(t)x^\ast(t))\right]\frac{\d t}{t},
$$
where the supremum is taken over all countably valued step functions
$[0,\infty[\ni t\mapsto x(t)\in{\cal O}$ vanishing in a neighborhood 
of zero and satisfying $x(t)+y(t)=\one$. In particular, if $\zeta_1(\one)=\zeta_2(\one)$, 
then $S(\zeta_1|\zeta_2)\geq 0$ with equality iff $\zeta_1=\zeta_2$.

Any unitary element $U\in\cO$ induces a $\ast$-automorphism
$$
A\mapsto\alpha_U(A)=U^\ast A U,
$$
and hence a state transformation $\omega\mapsto\omega\circ\alpha_U$.
Set $\eta=\one\otimes\nu_\i$. With this setup the entropy balance 
equation of~\cite{JP3,JP7,Pi1,PW1} reads as:  

\bet
Suppose that $U\in\Dom(\delta_\cR)$. Then
\beq 
S(\omega\circ\alpha_U|\eta)=S(\omega_\i|\eta)-\i\beta \omega_\i(U^\ast\delta_\cR(U)).
\label{ep-form}
\eeq
\eet

Denote by $\rho_U$ and $\nu_U$ the restriction of the transformed state 
$\omega\circ\alpha_U$ to $\cO_\cS$ and $\cO_\cR$ (i.e.,
$\rho_{U}(A)=\omega\circ\alpha_U(A\otimes\one)$ and 
$\nu_U(B)=\omega\circ\alpha_U(\one\otimes B)$ 
for $A\in\cO_\cS$ and $B\in\cO_\cR$). If $\cR$ is confined, then 
$$
-\i\omega_\i(U^\ast\delta_\cR(U))=-\i\omega_\i(U^\ast\i[H_\cR,U])
=\omega_\i(\alpha_U(H_\cR)-H_\cR)=\tr(\nu_UH_\cR)-\tr(\nu_\i H_\cR),
$$
and~\eqref{ep-form} reduces to~\eqref{akin-2}. 

For any states $\rho$ on $\cO_\cS$ and $\omega$ on $\cO$, Araki's perturbation 
formula for the relative entropy \cite{Ar1} (see also Proposition~6.2.32 
in~\cite{BR2} and Appendix~A of~\cite{Don}) gives
\beq\label{ArakiForm}
S(\omega|\rho\otimes\nu_\i)=S(\omega|\eta)-\omega(\log\rho).
\eeq
Setting $\omega=\omega_\i=\rho_\i\otimes\nu_\i$, this implies in particular that
\beq\label{Srhoform}
S(\rho_\i)=-S(\omega_\i|\eta).
\eeq

The entropy balance equation~\eqref{ep-form} allows for an analysis of 
Landauer's principle in the general setup of quantum statistical mechanics. The 
decrease in entropy of $\cS$ and the increase in energy of $\cR$ in the 
transition process $\omega_\i\rightarrow \omega\circ\alpha_U$ are 
$$
\Delta S= S(\rho_\i)-S(\rho_U),\qquad
\Delta\cQ=-\i\omega_\i(U^\ast\delta_\cR(U)).
$$
Writing~\eqref{ep-form} as 
\beq\label{barbados}
\Delta S +\sigma =\beta \Delta\cQ,
\eeq
and taking~\eqref{Srhoform} into account yields
\[
\sigma= S(\omega\circ\alpha_U|\eta)+S(\rho_U).
\]
Since $S(\rho_U)=-\omega\circ\alpha_U(\log\rho_U)$, Eq.\;\eqref{ArakiForm} further gives
\beq\label{sigmaOneForm}
\sigma=S(\omega\circ\alpha_U|\rho_U\otimes \nu_\i),
\eeq
and hence
\[
\sigma\geq0
\]
with equality iff $\omega\circ\alpha_U=\rho_U\otimes\nu_\i$. This implies the 
Landauer bound 
\[
\beta\Delta\cQ\geq\Delta S.
\]
for the state transformation induced by the inner $\ast$-automorphism $\alpha_U$.
This also settles the  conjecture of~\cite{ReWo} which has in fact been been known 
for many years. 

The analysis of the saturation of the Landauer bound is more delicate than in 
the case of a confined reservoir. It relies on the spectral analysis of modular
operators. We shall give one result in this direction.

\bep\label{Saturation}
Assume that the point spectrum of the standard Liouvillean $L_\cR$ is finite.
Then $\Delta S=\beta\Delta\cQ$ if and only if $\Delta S=\Delta\cQ=0$ in which 
case $\rho_U$ is unitarily equivalent to $\rho$ and $\nu_U=\nu_\i$
\eep

{\bf Remark.} If $\cR$ is confined, then the spectrum of $L_\cR$ is discrete and 
finite so that the above proposition applies. It also applies to the physically 
important class of ergodic extended reservoirs. Indeed, it follows 
from Theorem~1.2 in \cite{JP4} that $0$ is the only eigenvalue of $L_\cR$ if
$\nu_\i$ is an ergodic state for $(\cO_\cR,\tau_\cR)$.
It is an interesting  structural question  to characterize all reservoir systems for 
which the conclusions of Proposition~\ref{Saturation} holds.

One can continue with the abstract analysis of the Landauer principle in the 
above framework. As in the finite dimensional case, the bound
\beq
\sigma\geq\frac12\|\omega\circ\alpha_U-\rho_U\otimes\nu_\i\|^2
\label{lungs}
\eeq
follows from Eq.\;\eqref{sigmaOneForm}, the norm on the right hand side
being dual to the $C^\ast$-norm of $\cO$. Since
\begin{align*}
\|\nu_U-\nu_\i\|&=\sup_{0\not=A\in\cO_\cR}
\frac{|\omega\circ\alpha_U(\one\otimes A)-\rho_U\otimes\nu_\i(\one\otimes A)|}
{\|\one\otimes A\|}\\[4pt]
&\le\sup_{0\not=A\in\cO}\frac{|\omega\circ\alpha_U(A)
-\rho_U\otimes\nu_i(A)|}{\|A\|}
=\|\omega\circ\alpha_U-\rho_U\otimes\nu_\i\|,
\end{align*}
(\ref{lungs}) gives 
\beq\label{maiden}
\sigma\ge\frac12\|\nu_U-\nu_\i\|^2.
\eeq

Suppose that $\rho_\i>0$ and let $\rho_\f>0$ be a target state. Set 
$\omega_\f=\rho_\f\otimes\nu_\i$. Another application of Araki's 
perturbation formula gives 
\beq\label{ever}
\sigma=S(\omega\circ\alpha_U|\rho_U\otimes\nu_\i)
=S(\omega\circ\alpha_U|\omega_\f)-S(\rho_{U}|\rho_\f).
\eeq
Assume that there exists a sequence $U_n$ of unitary elements of $\cO$ such that $U_n\in\Dom(\delta_\cR)$ and
\beq\label{tra-wed}
\lim_{n\to\infty}\omega\circ\alpha_{U_n}(A)=\omega_\f(A)
\eeq
for all $A\in\cO$.
Since this implies that $\rho_{U_n}\to\rho_\f$, it follows from the 
entropy balance relation~\eqref{barbados} that 
\[
\liminf_{n\to\infty}\beta\Delta\cQ_n
=\liminf_{n\to\infty}\sigma_n+S(\rho_\i)-S(\rho_\f)
\geq S(\rho_\i)-S(\rho_\f).
\]
Moreover, 
\[
\lim_{n\to\infty}\beta\Delta\cQ_n=S(\rho_\i)-S(\rho_\f)
\]
if and only if
\[
\lim_{n\to\infty}\sigma_n=0,
\]
which, by~\eqref{ever}, is equivalent to
\beq\label{tra-quasistatic} 
\lim_{n\to\infty}S(\omega\circ\alpha_{U_n}|\omega_\f)= 0.
\eeq
The relation~\eqref{tra-quasistatic} quantifies the notion of quasi-static 
transition process. If~\eqref{tra-quasistatic} holds, then 
Inequality~\eqref{maiden} implies
\beq\label{meet}
\lim_{n\to\infty}\|\omega\circ\alpha_{U_n}-\omega_\f\|=0.
\eeq
On the other hand, the  norm convergence~\eqref{meet} does not 
imply~(\ref{tra-quasistatic}). Sufficient conditions for~\eqref{tra-quasistatic} 
are discussed in the foundational papers~\cite{Ar1, Ar2}. For example, if in 
addition to~\eqref{meet} there is $\lambda>0$ such that
\beq\label{improve}
\lambda\omega\circ\alpha_{U_n}\geq\omega_\f
\eeq
for all $n$, then~\eqref{tra-quasistatic} holds. A sufficient condition 
for~\eqref{improve} is that 
\[
\sup_{n}\|\e^{-\i\delta_\cR/2\beta}(U_n)\|<\infty.
\]

{\noindent\bf Remark.} If the quantum dynamical system $(\cO_\cR,\tau_\cR,\nu_\i)$ describes an infinitely extended reservoir in
thermal equilibrium at positive temperature and in a pure phase, then on physical grounds it is natural to  assume 
that the enveloping von Neumann algebra
$\pi_{\nu_\i}(\cO_\cR)''$ is an injective factor of type ${\rm III}_1$
(see, e.g., \cite{Ar4,ArW,Hu}). In this case, it is a simple consequence of a 
result of Connes and St\o rmer (Theorem 4 in~\cite{CSt}) and Kaplansky's density 
theorem (Corollary 5.3.7 in \cite{KR}) that there is a sequence 
of unitaries $U_n\in\Dom(\delta_\cR)$ such that~\eqref{meet} and 
hence~\eqref{tra-wed} holds.

\bigskip
Although one can go quite far by continuing the above structural analysis of the 
Landauer principle, we shall not pursue this direction  further. Instead, 
we shall focus on physically relevant realizations of $\alpha_U$'s by 
considering the dynamics of the coupled system $\cS+\cR$ and we shall 
analyze the Landauer principle in this context. Non-trivial dynamics are
characterized by interactions that allow energy/entropy flow between $\cS$ 
and $\cR$. We shall distinguish between instantaneously and adiabatically 
switched interactions.

\section{Instantaneously switched interactions}
\label{sec-isi}

\subsection{Setup}
For $K\in\cO^\sa$, the $\ast$-derivation
$$
\delta_K=\delta_\cR+\i[K,\,\cdot\,]
$$
generates a strongly continuous group $\tau_K^t=\e^{t\delta_K}$ of 
$\ast$-automorphisms of $\cO$. Self-adjoint elements of $\cO$ are
called local perturbations and the group $\tau_K$ is the local perturbation of 
$\tau_\cR$ induced by $K$. For example, if $H_\cS$ is the Hamiltonian of $\cS$ 
and $V$ describes the interaction of $\cS$ with $\cR$, then the dynamics of the
interacting system $\cS+\cR$ is given by $\tau_K$, with $K=H_\cS+V$.
In this section, we investigate the Landauer principle for the
dynamical system $(\cO,\tau_K)$.

The interacting dynamics can be expressed as
\[
\tau_K^t(A)=\tau_\cR^t(U^\ast_K(t)A U_K(t)),
\]
where the interaction picture propagator $U_K(t)$ is a family of unitary
elements of $\cO$ satisfying 
\begin{equation}\label{CocycleODE}
\i\partial_t U_K(t)
=U_K(t)\tau_\cR^{-t}(K),\qquad U_K(0)=\one.
\end{equation}
Hence, we have 
\[
\omega_\i\circ\tau_K^t=\omega_\i\circ\alpha_{U_K(t)},
\]
and we can apply the results of the previous section.
Assuming $K\in\Dom(\delta_\cR)$, it follows from the Dyson expansion
$$
U_K(t)=\one+\sum_{n=1}^\infty(-\i)^n
\int\limits_{0\le s_1\le\cdots\le s_n\le t}
\tau_\cR^{-s_1}(K)\cdots\tau_\cR^{-s_n}(K)\d s_1\cdots\d s_n
$$
that $U_K(t)\in\Dom(\delta_\cR)$ for all $t\in\rr$ and 
Eq.\;\eqref{barbados} gives
\beq\label{barbados-no}
\Delta S(K,t)+\sigma(K,t)=\beta \Delta\cQ(K,t),
\eeq
where 
\[
\Delta\cQ(K,t)
=-\i\omega_\i(U_K^\ast(t)\delta_\cR(U_K(t))),
\] 
and
\[
\Delta S(K,t)=S(\rho_\i)-S(\rho_K(t)),\qquad
\sigma(K,t)=S(\omega_\i\circ\tau_K^t|\rho_K(t)\otimes\nu_\i),
\]
$\rho_K(t)$ denoting the restriction of $\omega_\i\circ\tau_K^t$ 
to $\cO_\cS$. 

\bigskip
{\noindent\bf Remark.} One easily checks that 
$T(t)=\i U^\ast_K(t)\delta_\cR(U_K(t))+\tau_\cR^{-t}(K)$
satisfies the Cauchy problem
$$
\partial_t T(t)=\i[\tau_\cR^{-t}(K),T(t)],\qquad T(0)=K.
$$
Comparing with Eq.\;\eqref{CocycleODE}, we infer
$T(t)=U_K^\ast(t)K U_K(t)$, so that
$$
-\i U^\ast_{K}(t)\delta_\cR(U_{K}(t))
=\tau_\cR^{-t}(K-\tau_{K}^t(K)),
$$
and therefore
$$
\Delta\cQ(K,t)=\omega_\i(K-\tau_{K}^t(K)).
$$
Conservation of the total energy leads to the conclusion that $\Delta\cQ(K,t)$ is 
indeed the change of the reservoir energy. Since
$$
\partial_t(K-\tau_{K}^t(K))
=\tau_{K}^t(-\delta_\cR(K)),
$$
one  can  further write
\[
\Delta\cQ(K,t)=-\int_0^t\omega_\i(\tau_{K}^s(\Phi))\d s,
\]
where 
\[
\Phi=\delta_\cR(K),
\]
is the observable describing the instantaneous energy flux out of $\cR$.

\subsection{The Landauer principle in the  large time limit}
\label{sec-inst}
We shall now consider realizations of the  state transition 
$\rho_\i\to\rho_\f$ and the corresponding entropic balance as a limiting case
of the transition $\rho_\i\to\rho_K(t)$ as $t\to\infty$. To simplify the
discussion, we shall assume here and in the following that the equilibrium state 
$\nu_\i$ describes a pure thermodynamic phase of $\cR$, i.e., that it is an 
extremal 
$(\tau_\cR,\beta)$-KMS state. This implies that for $K\in\cO^\sa$ there is
a unique $(\tau_K,\beta)$-KMS state in $\cN_{\omega_\i}$ which we denote
by $\mu_K$. Let $\varrho_K$ be its restriction to $\cO_\cS$. We observe 
that $\cN_{\mu_K}=\cN_{\omega_\i}$. 

The following proposition shows that by an appropriate choice of $K$
we can reach any faithful\footnote{The cases where the  
target state is not faithful are handled by an additional limiting argument 
that we will describe later.} target state of $\cO_\cS$ with the 
$(\tau_K,\beta)$-KMS state $\mu_K$.

\bep\label{HTarget}
Let $\rho_\f>0$ be a state on $\cO_\cS$ and $V\in\cO^\sa$. Then there exists 
$\delta>0$ and a real analytic function 
$]-\delta,\delta[\ni\lambda\mapsto H_\lambda\in\cO_\cS^\sa$ such that $H_0=-\beta^{-1}\log\rho_\f$ 
and $\varrho_{K_\lambda}=\rho_\f$ for $K_\lambda=H_\lambda+\lambda V$ and any $\lambda\in ]-\delta, \delta[$.
\eep

Our main dynamical assumption is:

\begin{quote} {\bf Assumption A.} There exists  $\gamma\in ]-\delta,\delta[$ such that the KMS state $\mu_{K_\gamma}$ is mixing for the 
dynamical system $(\cO,\tau_{K_\gamma})$.
\end{quote}

We now explore the  consequences of this assumption on the long time asymptotics
of entropy balance (note that obviously $\gamma\not=0$). The first is 
$$
\lim_{t\to\infty}\rho_{K_\gamma}(t)=\rho_\f.
$$
Furthermore, 
\beq\label{rainysat}
\begin{split}
\Delta S&=\lim_{t\to\infty}\Delta S(K_\gamma,t)
=S(\rho_\i)-S(\rho_\f),\\[2mm]
\Delta\cQ(\gamma)&=\lim_{t\to\infty}\Delta\cQ(K_\gamma,t)
=\omega_\i(K_\gamma)-\mu_{K_\gamma}(K_\gamma).
\end{split}
\eeq

It follows from~\eqref{barbados-no} that 
\[
\sigma(\gamma)=\lim_{t\rightarrow \infty}\sigma(K_\gamma, t)
\]
also exists and that 
\beq\label{dog}
\Delta S+\sigma(\gamma)=\beta\Delta\cQ(\gamma).
\eeq
Clearly, $\sigma(\gamma)\geq 0$, and the relation (\ref{dog}) gives the Landauer principle for the transition process 
$\rho_\i\rightarrow\rho_\f$ realized  by the large time limit 
$t\rightarrow \infty$. 

One does not expect that  the Landauer bound can be saturated by an  instantaneously 
switched   interaction and that is indeed the case.
\bep 
\beq\label{strict-po}
\sigma(\gamma)>0.
\eeq
\label{roof}
\eep

This completes our analysis of the Landauer principle for instantaneously switched 
interactions.

\bigskip
{\noindent\bf Remark 1.} The above analysis  extends with no changes to $W^\ast$-dynamical 
systems. Unbounded interactions $V$ satisfying the general assumptions of~\cite{DJP} 
are also allowed.

\medskip

{\noindent\bf Remark 2.} In the Landauer erasing principle, $\rho_\i=\one/d$ 
and $\rho_\f=|\psi\rangle\langle\psi|$. Pure target states are 
thermodynamically singular and cannot be directly reached by the action of a 
thermal reservoir unless the reservoir is at zero temperature. The proper way to 
formulate the Landauer principle for pure states is to examine the stability of 
the entropy balance equation of the processes with faithful target states
$\rho_\f^\prime$ in a vicinity of $\rho_\f$. For instantaneously 
switched interactions there is no stability. As 
$\rho_\f^\prime\to\rho_\f$, $S(\rho_\f^\prime)\to S(\rho_\f)=0$. 
However, in this limit $\sigma(\gamma)\to\infty$ and 
$\Delta\cQ(\gamma)\to\infty$. This singularity is due to an
instantaneous change of the Hamiltonian.
As we shall see in the next section, if the change of the Hamiltonian is 
adiabatic, this singularity is 
absent.

\medskip

\noindent{\bf Remark 3.} It follows from 
Araki's perturbation theory of KMS states that the map 
\[ ]-\delta, \delta[ \ni \lambda\mapsto\Delta\cQ(\lambda)= \omega_\i(K_\lambda)-\mu_{K_\lambda}(K_\lambda)\]
is  real analytic and that 
$$
\Delta\cQ(0)=\rho_\i(H_0)-\rho_\f(H_0).
$$
The relation (\ref{dog}) defines $\sigma(\lambda)$ for 
$\lambda \in ]-\delta, \delta[$ and 
\[
\sigma(0)=S(\rho_\i|\rho_\f).
\]

\medskip

\medskip
{\noindent\bf Remark 4.} For many models, Assumption {\bf A} is satisfied in a stronger form: 

\begin{quote} {\bf Assumption A'.} There exists $\lambda_0>0$ such that for $0 <|\lambda|<\lambda_0$  the KMS state $\mu_{K_\lambda}$ is mixing for the 
dynamical system $(\cO,\tau_{K_\lambda})$.
\end{quote}

In this case the entropy balance equation 
\beq \Delta S + \sigma(0)=\beta \Delta {\cal Q}(0)
\label{clouds}
\eeq
gives the Landauer principle for the transition process $\rho_\i \rightarrow \rho_\f$ realized by the double limit $t\rightarrow \infty$, 
$\lambda\rightarrow 0$. The relation (\ref{clouds}) is certainly 
expected in view  of the Lebowitz-Spohn weak coupling 
limit thermodynamics of open quantum systems~\cite{LeSp,JPW}. Under suitable 
assumptions, in the van Hove scaling limit $\lambda\rightarrow0$,
$t\rightarrow\infty$ with $\bar t=\lambda^2 t$ fixed, the reduced dynamics of 
$\cS$ is described by a quantum dynamical semigroup on $\cO_\cS$, 
\[T_{\bar t}(A)=\e^{\bar t {\cal K}}(A),\]
where $\cK$ is the so-called {\em Davies generator} in the Heisenberg picture. 
Under the usual effective coupling assumptions one has\footnote{The adjoint is 
taken with respect to the inner product $\langle A,B\rangle=\tr(A^\ast B)$ on $\cO_\cS$}
\[
\lim_{\bar t\to\infty}\e^{\bar t\cK^\dagger}(\rho)=\rho_\f
\]
for any state $\rho$ on $\cO_\cS$.
This relation defines the transition process $\rho_\i\to\rho_\f$ in the van Hove
scaling limit. The Lebowitz-Spohn entropy balance equation is 
\[
S(\rho_\i)-S(\e^{\bar t\cK^\dagger}(\rho_\i))
+S(\rho_\i|\e^{\bar t\cK^\dagger}(\rho_\i))=\beta\Delta\bar\cQ(\bar t),
\]
where 
\[
\Delta\bar\cQ(\bar t)=\rho_\i(\e^{\bar t\cK}(H_0)- H_0).
\]
It follows that 
\[
\Delta\bar\cQ=\lim_{{\bar t}\to\infty}\Delta\bar\cQ(\bar t)
=\rho_\f(H_0)-\rho_\i(H_0),
\]
and one derives
\beq
\beta\Delta\bar\cQ=S(\rho_\i|\rho_\f)+S(\rho_\i)-S(\rho_\f).
\label{corrections1}
\eeq
Since the van Hove weak coupling limit is expected to yield the first 
non-trivial contribution (in the coupling constant $\lambda$) to the microscopic 
thermodynamics, the identity~\eqref{clouds}=\eqref{corrections1} is 
certainly not surprising. A somewhat surprising fact is that Assumption {\bf A'} 
is only vaguely related to the assumptions of the weak coupling limit theory 
\cite{Dav,DF,LeSp}.

\medskip

{\noindent\bf Remark 5.}  Specific physically relevant models (spin-boson model, 
spin-fermion model, electronic black box model, locally interacting fermionic 
systems) for which Assumption {\bf A'} holds   are discussed in 
\cite{AM,AJPP1,AJPP2,BFS,BM,dRK,DJ,FMU,FMSU,JOP1,JOP2,JP1,JP6,MMS1,MMS2}.

\section{Adiabatically switched interactions}
\label{sec-asi}

Our next topic is the optimality of the Landauer bound in the context of time 
dependent Hamiltonian dynamics of $\cS+\cR$. We shall assume that the reader is 
familiar with basic results concerning non-autonomous perturbations of 
$C^\ast$-dynamical systems (see Section~5.4.4 in~\cite{BR2} and the Appendix 
to Section~IV.5 in~\cite{Si}). 

\subsection{Setup}

Let $K:[0,1]\to\cO^\sa\cap\Dom(\delta_\cR)$ be a continuous function which we 
assume to be twice continuously differentiable on $]0,1[$ with uniformly bounded
first and second derivatives. For $T>0$, we define the rescaled function $K_T$ by
$$
K_T(t)=K(t/T).
$$
Let $[0,T]\ni t\mapsto\alpha_{K_T}^t$ be the non-autonomous $C^\ast$-dynamics 
defined by the Cauchy problem
\beq\label{AlphaDef}
\partial_t \alpha_{K_T}^t(A)
=\alpha_{K_T}^t(\delta_\cR(A)+\i[K_T(t),A]),\qquad\alpha_{K_T}^0(A)=A.
\eeq
We recall that $\{\alpha_{K_T}^t\}_{t\in[0,T]}$ is a strongly continuous family  
of $\ast$-automorphisms of $\cO$ given by
$$
\alpha_{K_T}^t(A)=\tau_\cR^t(A) +\sum_{n=1}^\infty\,
\int\limits_{0\le s_1\le\cdots\le s_n\le t}
\i[\tau_\cR^{s_1}(K_T(s_1)),\i[
\cdots,\i[\tau_\cR^{s_n}(K_T(s_n)),\tau_\cR^t(A)]]]
\d s_1\cdots\d s_n.
$$
Moreover, the interaction representation
\beq\label{InterGamma}
\tau_\cR^{-t}\circ\alpha_{K_T}^t(A)
=\tau_\cR^{-t}(\Gamma_{K_T}(t))^\ast A\tau_\cR^{-t}(\Gamma_{K_T}(t)),
\eeq
holds with a family of unitaries $\Gamma_{K_T}(t)\in\cO$ satisfying the
Cauchy problem
\beq\label{GammaK}
\i\partial_t\Gamma_{K_T}(t)=\tau_\cR^t(K_T(t))\Gamma_{K_T}(t),\qquad
\Gamma_{K_T}(0)=\one.
\eeq

We denote by $\rho_T$ the restriction of $\omega_\i\circ\alpha_{K_T}^T$ to
$\cO_\cS$. Our assumptions ensure that $\Gamma_{K_T}(T)\in\Dom(\delta_\cR)$ and
Eq.\;\eqref{barbados} gives
\beq\label{ep-ad}
\Delta S_T+\sigma_T=\beta\Delta\cQ_T,
\eeq
where
$$
\Delta S_T=S(\rho_\i)-S(\rho_T),\qquad
\Delta\cQ_T
=-\i\omega_\i\left(\Gamma_{K_T}^\ast(T)\delta_\cR(\Gamma_{K_T}(T))\right),
$$
and
$$
\sigma_T=S(\omega_\i\circ\alpha_{K_T}^T|\rho_T\otimes\nu_\i).
$$

To interpret the right hand side of Eq.\;\eqref{ep-ad}, we write
$$
\Delta\cQ_T=Q(T)-Q(0)=\int_0^T\partial_t Q(t)\d t,
$$ with 
$Q(t)=-\i\omega_\i(\Gamma_{K_T}^\ast(t)\delta_\cR(\Gamma_{K_T}(t)))$. It follows
from the differential equation~\eqref{GammaK} that
$$
\partial_t Q(t)
=-\omega_\i\left(
\Gamma_{K_T}^\ast(t)\tau_\cR^t(\delta_\cR(K_T(t)))\Gamma_{K_T}(t)\right),
$$
and Eq.\;\eqref{AlphaDef}-\eqref{InterGamma} give
\[
\begin{split}
\partial_t Q(t)&=-\omega_\i\circ\alpha_{K_T}^t\left(\delta_\cR(K_T(t))\right)\\[3mm]
&=-\frac{\d\ }{\d t}\,\omega_\i\circ\alpha_{K_T}^t(K_T(t))
+\omega_\i\circ\alpha_{K_T}^t\left(\frac{\d\ }{\d t}\,K_T(t)\right).
\end{split}
\]
This leads to 
\beq\label{FirstLaw}
\Delta\cQ_T+\omega_\i\circ\alpha_{K_T}^T(K_T(T))-\omega_\i(K_T(0))
=\int_0^T\omega_\i\circ\alpha_{K_T}^t\left(P_T(t)\right)\d t,
\eeq
where $P_T(t)=\partial_t K_T(t)$ is the instantaneous power injected into the
system $\cS+\cR$. Energy conservation yields that $\Delta\cQ_T$ is the 
total change in the energy of the subsystem $\cR$ from time $t=0$ to time $t=T$.

\subsection{The Landauer principle in the adiabatic limit}
\label{AL}

We shall now consider the adiabatic limit $T\to\infty$. Our main assumption in 
this section concerns the instantaneous $C^\ast$-dynamics
$\tau_{K(\gamma)}$.

\begin{quote}{\bf  Assumption B.} For $0<\gamma<1$, the
$(\tau_{K(\gamma)},\beta)$-KMS state 
$\mu_{K(\gamma)}$ is ergodic for the dynamical system $(\cO,\tau_{K(\gamma)})$.
\end{quote}

The Avron-Elgart adiabatic theorem~\cite{AE,Teu} and Araki's perturbation theory 
of KMS states give (\cite{ASF1}-\cite{ASF3}, \cite{JP8}):

\bet\label{adiabatic}
Suppose that Assumption {\bf B} holds. Then one has
\[
\lim_{T\to\infty}\|\mu_{K(0)}\circ\alpha_{K_T}^{\gamma T}-\mu_{K(\gamma)}\|=0
\]
for all $\gamma\in[0,1]$.
\eet
For completeness and the reader's convenience the proof of Theorem \ref{adiabatic} 
is given in Section \ref{sec-proof}.

Let $\rho_\f$ be a given faithful target state of $\cS$ and set
\[
\omega_\f=\rho_\f\otimes\nu_\i.
\]

According to Theorem~\ref{adiabatic}, to achieve the transition 
$\rho_\i\to\rho_\f$ in the limit $T\to\infty$,  it suffices to assume that Assumption 
{\bf B} holds for $K(\gamma)$ satisfying the boundary conditions
\[
K(0)=-\beta^{-1}\log \rho_\i,\qquad
K(1)=-\beta^{-1}\log \rho_\f.
\]
Indeed, these conditions ensure that $\mu_{K(0)}=\omega_\i$ and $\mu_{K(1)}=\omega_\f$ so that
$$
\lim_{T\to\infty}\omega_\i\circ\alpha_{K_T}^T
=\lim_{T\to\infty}\mu_{K(0)}\circ\alpha_{K_T}^T
=\mu_{K(1)}=\omega_\f.
$$
Theorem~\ref{adiabatic} further implies that 
\[
\Delta S=\lim_{T\to\infty}\Delta S_T=S(\rho_\i)-S(\rho_\f).
\]
Moreover, rewriting Eq.\;\eqref{FirstLaw} as
$$
\Delta\cQ_T=\int_0^1\omega_\i\circ\alpha_{K_T}^{\gamma T}\left(
\partial_\gamma K(\gamma)\right)\d\gamma
-\beta^{-1}\omega_\i\circ\alpha_{K_T}^T(\log\rho_\f)+\beta^{-1}\omega_\i(\log\rho_\i),
$$
we get
$$
\Delta\cQ=\lim_{T\rightarrow \infty}\Delta\cQ_T
=\int_0^1\mu_{K(\gamma)}(\partial_\gamma K(\gamma))\d\gamma
+\beta^{-1}\Delta S.
$$
The balance equation~\eqref{ep-ad} yields that
\[
\Delta S +\sigma=\beta \Delta\cQ
\]
with
\[
\sigma=\lim_{T\to\infty}\sigma_T
=\beta\int_0^1\mu_{K(\gamma)}(\partial_\gamma K(\gamma))\d\gamma.
\]
Clearly, $\sigma \geq 0$. The adiabatic limit is a quasi-static process and one may expect the optimality of the Landauer bound. This is indeed the case. 
\bep\label{thm-td-1}
\[
\sigma=0.
\]
\eep
The proof of the last result  requires modular theory. Note however that for finite reservoirs  the relation 
\[
\int_{0}^1\mu_{K(\gamma)}(\partial_\gamma K(\gamma))\d\gamma=0
\]
is easily derived\footnote{On the other hand,  Theorem \ref{adiabatic} and relation $\lim_{T\rightarrow \infty}\sigma_T=\sigma$ {\em cannot} hold for finite reservoirs.}:
\begin{align*}
\int_{0}^1\mu_{K(\gamma)}(\partial_\gamma K(\gamma))\d\gamma
&=\int_{0}^1
\frac{\tr\left(\e^{-\beta(H_\cR+K(\gamma))}\partial_\gamma K(\gamma)\right)}%
{\tr\left(\e^{-\beta(H_\cR+K(\gamma))}\right)}\d\gamma\\[2mm]
&=-\frac1\beta\int_0^1\partial_\gamma\log\tr\left(
\e^{-\beta(H_\cR+K(\gamma))}\right)\d\gamma\\[2mm]
&=-\frac{1}{\beta}\left(\log\tr (\omega_\f)-\log\tr(\omega_\i)\right)=0.
\end{align*}
This completes our mathematical analysis of the Landauer principle for 
adiabatically switched interactions. 

\bigskip
\noindent{\bf Remark 1.} Regarding the remarks at the end of Section~\ref{sec-inst}, 
Remark 1 applies to the results of this section as well. In the adiabatic case 
the entropy production term vanishes and the instability discussed in
Remark 2 is absent. Remark 4 also extends to the adiabatic setting (see~\cite{DS} 
for the discussion of the adiabatic theorem and~\cite{AHHH} for a discussion of 
the Landauer principle in the van Hove weak coupling limit).
Since mixing implies ergodicity, the physically relevant models 
for which Assumption {\bf B} has been verified are listed in Remark 5. 

\medskip
\noindent{\bf Remark 2.} The Narnhoffer-Thirring adiabatic theorem of quantum 
statistical mechanics~\cite{NT} is based on $C^\ast$-scattering and requires 
$L^1$-asymptototic Abelianess which is stronger than our ergodicity assumption 
{\bf B}. The physically relevant models satisfying $L^1$-asymptotic Abelianess 
are discussed in~\cite{AM,AJPP1,AJPP2,BM,FMU,FMSU,JOP2}. If $L^1$-asymptotic 
Abelianess holds, then the Landauer principle for adiabatically switched
interactions can be further refined. The result of this analysis is given 
in~\cite{Han}.

\section{Discussion}
\label{sec-disc}
In this section we comment on the key ingredients involved in the analysis of the
Landauer principle presented in Sections~\ref{sec-ep-bal}--\ref{sec-asi}, and on 
their relation with the work~\cite{ReWo}. 

\bigskip
\noindent{\em The entropy balance equation.} 
Relation~\eqref{ep-form} is  a model-independent structural identity linked to 
the KMS condition and modular theory. It is tautological in the finite dimensional 
case. The general case follows from Araki's perturbation theory of the 
modular structure. The mathematical analysis of the second law of thermodynamics
starts with the entropy balance equation but certainly does not end
there\footnote{In the literature, the entropy balance equation~\eqref{ep-form} is 
sometimes called "finite time second law of thermodynamics" reflecting the fact 
that in typical application $U$ is a unitary cocycle describing time evolution 
over a finite time period.}\cite{ASF1,ASF2,ASF3}. The thermodynamic behavior of the 
coupled system $\cS+\cR$ emerges only in the thermodynamic limit in which the reservoir 
$\cR$ becomes infinitely extended. In the large time limit the coupled system settles 
into a steady state, substantiating the zeroth law of thermodynamics 
\cite{BFS,DJ,FM,JP1}.

These two limiting processes, large reservoir size and large time, have been pillars 
of the mathematical theory of open quantum systems since its 
foundations~\cite{Rob,BR1,BR2}. In a sense, the same applies to the Landauer 
principle and this is the main message of this note: the control of the entropy
balance equation for open quantum systems with infinitely extended reservoirs
in the large time (or adiabatic) limit is one of the central issues in the analysis 
of the Landauer principle within quantum statistical mechanics. This brings us to our 
second point.

\bigskip
\noindent{\em Confined reservoirs.} 
A typical physical example of a confined reservoir is a Fermi gas or a Bose gas 
in thermal equilibrium confined to a finite box. Confined reservoirs are not ergodic
and lead to quasi-periodic dynamics when coupled to a finite system $\cS$.
The analysis of the large time asymptotics of such systems requires some
time averaging which is not compatible with the formulation of Landauer's
principle. In this context, 
one may say that the main contribution of~\cite{ReWo} concerns estimates 
regarding the accuracy of the Landauer principle for confined reservoirs. 

\bigskip
\noindent{\em Ergodicity.}
The large time asymptotics of the microscopic system $\cS$ coupled to the thermal
reservoir $\cR$ is critically linked to the ergodic properties  (Assumptions {\bf A} 
and {\bf B}) of the dynamical system which describes the joint system $\cS+\cR$ in the 
framework of statistical mechanics.
As we have shown, ergodicity allows for arbitrary transition $\rho_\i\to\rho_\f$
of the system $\cS$ in the adiabatic limit with the minimal energy dissipation 
predicted by Landauer.
Needless to say, Assumptions {\bf A} and {\bf B}, which are part of the zeroth law of 
thermodynamics,  are notoriously difficult to prove for physically relevant 
models. In particular, they cannot hold in the framework of~\cite{ReWo}, where the 
reservoirs are confined.





\bigskip

\noindent{\em Conclusion.} The claim of the authors in \cite{ReWo} that they 
have proven the Landauer principle in quantum statistical mechanics may lead to a 
confusion regarding some foundational aspects of mathematical theory of open 
quantum systems and we have attempted to clarify this point. The complementary 
analysis of the Landauer principle presented in 
Sections~\ref{sec-ep-bal}--\ref{sec-asi} relies on the entropy balance equation, 
Araki's perturbation theory of KMS states, and the Avron-Elgart adiabatic 
theorem. It is a simple consequence of well-known and deep structural results. 
The workers in quantum information theory appear unaware of this fact. 
From the 
point of view of state-of-the-art quantum statistical mechanics, the 
interesting aspect of the Landauer principle concerns the verifications of 
Assumptions {\bf A} and {\bf B}. The models for which this has been achieved are discussed in \cite{AM,AJPP1,AJPP2,BFS,BM,dRK,DJ,FMU,FMSU,JOP1,JOP2,JP1,JP6,MMS1,MMS2}. One may say that one  of the main challenges of quantum statistical mechanics at 
the moment is to extend  the class of physically relevant models for which
Assumptions {\bf A} and {\bf B} can be proved. The progress in this direction 
requires novel ideas and techniques in the study of the large time dynamics of 
infinitely extended Hamiltonian quantum statistical models.

\section{Proofs}
\label{sec-proof}

\paragraph{Preliminaries.}
We start with some general properties of the GNS representation
$(\fH,\pi,\Omega)$ of $\cO$ associated to the state 
$\omega_\i=\rho_\i\otimes\nu_\i$. This material is standard and we refer the 
reader to \cite{BR1, BR2,DJP} for a detailed exposition and proofs. We denote by 
$\fM=\pi(\cO)^{\prime\prime}$ the enveloping von Neumann algebra and by 
$\cP\subset\fH$ and 
$J$ the natural cone and modular conjugation of the pair $(\fM,\Omega)$.
Any state $\omega\in\cN_{\omega_\i}$ has a unique standard representative,
a unit vector $\Psi\in\cP$ such that $\omega(A)=\langle \Psi,\pi(A)\Psi\rangle$ for all 
$A\in\cO$. The standard Liouvillean of a strongly continuous group $\varsigma$ of 
$\ast$-automorphisms of $\cO$ is the unique self-adjoint operator $L$ on $\fH$ 
such that
$$
\pi(\varsigma^t(A))=\e^{\i tL}\pi(A)\e^{-\i tL},\qquad\e^{\i tL}\cP\subset\cP,
$$
for all $t\in\rr$ and all $A\in\cO$.

Let $L_0$ be the standard Liouvillean of a group $\varsigma_0^t=\e^{t\delta_0}$
of $\ast$-automorphisms of $\cO$ and $\Phi_0$ the standard representative of a 
$(\varsigma_0,\beta)$-KMS state $\omega_0\in\cN_{\omega_\i}$.
If $Q\in\cO^\sa$ and $\delta_Q=\delta_0+\i[Q,\,\cdot\,]$, then the standard 
Liouvillean of the locally perturbed group $\varsigma_Q^t=\e^{t\delta_Q}$ is 
$$
L_Q=L_0+\pi(Q)-J\pi(Q)J.
$$
Moreover, $\Phi_0\in\Dom(\e^{-\beta(L_0+\pi(Q))/2})$ and the vector
$$
\Psi_Q=\frac{\Phi_Q}{\|\Phi_Q\|},\qquad
\Phi_Q=\e^{-\beta(L_0+\pi(Q))/2}\Phi_0,
$$
is the standard representative of a $(\varsigma_Q,\beta)$-KMS state. In particular, 
one has $\Psi_Q\in\Ker(L_Q)$.

We shall need the following perturbative expansion of the unnormalized 
KMS vector $\Phi_Q$. 
For any $Q_1,\ldots,$ $Q_n\in\fM$ and $(\beta_1,\ldots,\beta_n)\in T_{\beta,n}
=\left\{(\beta_1,\ldots,\beta_n)\in\rr_+^n,|\,
\beta_1+\cdots+\beta_n\leq\beta/2\right\}$ one has
\[
\Phi_0\in\Dom\left(\e^{-\beta_1L_0} Q_1\cdots\e^{-\beta_n L_0}Q_n\right).
\]
Moreover, the map 
\[
T_{\beta,n}\ni(\beta_1,\ldots,\beta_n)
\mapsto\e^{-\beta_1L_0} Q_1\cdots\e^{-\beta_n L_0}Q_n\Phi_0\in{\mathfrak H},
\]
is continuous and satisfies
\beq
\sup_{(\beta_1,\ldots,\beta_n)\in T_{\beta,n}}
\|\e^{-\beta_1L_0}Q_1\cdots\e^{-\beta_n L_0}Q_n\Phi_0\|
\leq \|Q_1\|\cdots\|Q_n\|.
\label{araki-est}
\eeq
The vector $\Phi_Q$ has the norm convergent expansion 
\beq
\Phi_Q=\sum_{n=0}^\infty (-1)^n \int_{T_{\beta, n}}
\e^{-\beta_1L_0}\pi(Q)\cdots\e^{-\beta_n L_0}
\pi(Q)\Phi_0\,\d\beta_1\cdots\d\beta_n.
\label{araki-exp}
\eeq

For $\delta Q\in\cO^\sa$, the following chain rule applies
$$
\Phi_{Q+\delta Q}=\e^{-\beta(L_Q+\pi(\delta Q))/2}\Phi_Q,
$$
(see Theorem 5.1 (6) in \cite{DJP}).
It follows from the expansion~\eqref{araki-exp} and the 
estimate~\eqref{araki-est} that the map $\cO^\sa\ni Q\mapsto\Phi_Q\in\fH$ is 
differentiable. Its derivative at $Q$ is the map
\beq\label{PhiPrime}
\Phi'_Q:\delta Q\mapsto-\int_0^{\beta/2}\e^{-sL_Q}\pi(\delta Q)\Phi_Q\d s.
\eeq
The same argument shows that if $\alpha\mapsto Q(\alpha)$ is a real analytic
function from some open subset of $\rr^n$ to $\cO^\sa$, then the function
$\alpha\mapsto\Phi_{Q(\alpha)}$ is also real analytic.

\bigskip
\paragraph{Proof of Proposition~\ref{Saturation}} To simplify the notation, we write 
$\rho=\rho_\i$, $\nu=\nu_\i$, $\omega=\omega_\i$. We denote by $\Omega_U\in\cP$ the
standard representative of the state $\omega_U$. 
The modular operator $\Delta_\omega$ and the relative modular operator 
$\Delta_{\omega_U|\omega}$ are positive operators on $\fH$ satisfying
$$
J\Delta_\omega^{1/2}\pi(A)\Omega=\pi(A)^\ast\Omega,\qquad
J\Delta_{\omega_U|\omega}^{1/2}\pi(A)\Omega=\pi(A)^\ast\Omega_U,
$$
for all $A\in\cO$. It follows from $\omega_U=\omega\circ\alpha_U$ that
 $\Omega_U=\pi(U)J\pi(U)J\Omega$. Since $J\pi(U)J\in\fM'$, one has
\begin{align*}
J\Delta_{\omega_U|\omega}^{1/2}\pi(A)\Omega&=\pi(A)^\ast\pi(U)J\pi(U)J\Omega\\
&=J\pi(U)J\pi(A)^\ast\pi(U)\Omega\\
&=J\pi(U)JJ\Delta_\omega^{1/2}\pi(U)^\ast\pi(A)\Omega,
\end{align*}
and the cyclic property of $\Omega$ allows us to conclude that
\beq
\Delta_{\omega_U|\omega}=\pi(U)\Delta_\omega\pi(U)^\ast.
\label{DeltaU}
\eeq

The product structure of the state $\omega$
induces the factorization $\fH=\fH_\cS\otimes\fH_\cR$ where $\fH_\cS=\cO_\cS$ equipped 
with the inner product $\langle X,Y\rangle=\tr(X^\ast Y)$ and the Hilbert
space $\fH_\cR$ carries a GNS representation of $\cO_\cR$ induced by the state $\nu$.
Moreover, one has
\beq
\Delta_{\omega}=\Delta_\rho\otimes\Delta_\nu,\qquad
\Delta_{\omega_U|\omega}=\Delta_{\rho_U|\rho}\otimes\Delta_{\nu},
\label{DeltaFact}
\eeq
where $\Delta_\rho$, $\Delta_{\nu}$ and $\Delta_{\rho_U|\rho}$ are respectively
the modular operator of the state $\rho$, the modular operator 
of the state $\nu$, and the relative modular operator of the state $\rho_U$ 
w.r.t.\;$\rho$. The operators $\Delta_\rho$ and $\Delta_{\rho_U|\rho}$ act on $\fH_\cS$
according to
\beq
\Delta_\rho X=\rho X\rho^{-1},\qquad\Delta_{\rho_U|\rho}X=\rho_U X\rho^{-1},
\label{DeltaForm}
\eeq
(see, e.g., Section~2.12 in \cite{JOPP}). In particular, they have discrete spectra.

Denote by $\Delta_{\omega,\rm p}$ and $\Delta_{\omega_U|\omega,\rm p}$ the pure point 
parts of $\Delta_\omega$ and $\Delta_{\omega_U|\omega}$. Eq.\;\eqref{DeltaFact} implies that
\beq
\Delta_{\omega,\rm p}=\Delta_\rho\otimes\Delta_{\nu,{\rm p}},\qquad
\Delta_{\omega_U|\omega,\rm p}=\Delta_{\rho_U|\rho}\otimes\Delta_{\nu,{\rm p}},
\label{DeltaP}
\eeq
where $\Delta_{\nu,{\rm p}}$ is the pure point part of $\Delta_\nu$.
Since $\Delta_\nu=\e^{-\beta L_\cR}$, the operators $\Delta_{\omega,\rm p}^{\i\alpha}$ 
and $\Delta_{\omega_U|\omega,\rm p}^{\i\alpha}$ are trace class by assumption and
it follows from Eq.\;\eqref{DeltaU} that these two operators are
unitarily equivalent so that
$$
\tr(\Delta_{\omega,\rm p}^{\i\alpha})=\tr(\Delta_{\omega_U|\omega,\rm p}^{\i\alpha}),
$$
for all $\alpha\in\cc$. Using Eq.\;\eqref{DeltaForm} and \eqref{DeltaP}, an
explicit calculation yields
$$
\tr(\Delta_{\omega,\rm p}^{\i\alpha})=
\tr(\Delta_\rho^{\i\alpha})\,\tr(\Delta_{\nu,{\rm p}}^{\i\alpha})
=\tr(\rho^{\i\alpha})\,\tr(\rho^{-\i\alpha})
\,\tr(\Delta_{\nu,{\rm p}}^{\i\alpha}),
$$
$$
\tr(\Delta_{\omega_U|\omega,\rm p}^{\i\alpha})
=\tr(\Delta_{\rho_U|\rho}^{\i\alpha})\,\tr(\Delta_{\nu,{\rm p}}^{\i\alpha})
=\tr(\rho_U^{\i\alpha})\tr(\rho^{-\i\alpha})\,\tr(\Delta_{\nu,{\rm p}}^{\i\alpha}).
$$
Thus, we conclude that
$$
\tr(\rho^{\i\alpha})=\tr(\rho_U^{\i\alpha}),
$$
for all $\alpha\in\cc$,
which implies that $\rho$ and $\rho_U$ are unitarily equivalent.
\hfill\qed

\bigskip

\paragraph{Proof of Proposition \ref{HTarget}} The proof is based on an application of  
the real analytic implicit function theorem.
Denote by $\mathfrak X$ the real vector space $\{X\in\cO_\cS^\sa\,|\,\tr(X)=0\}$
equipped with the inner product $( X,Y)=\tr(XY)$. Let 
$$
\rr\times\mathfrak{X}\ni(\lambda,X)\mapsto F(\lambda,X)
=\rho_{X+\lambda V}-\rho_\f\in\mathfrak{X}.
$$
First, note  that $F$ is real analytic. 
Moreover, for any $X\in\mathfrak{X}$, one has
$$
F(0,X)=\frac{\e^{-\beta X}}{\tr(\e^{-\beta X})}-\rho_\f,
$$
and  $F(0,X)=0$ iff $X=H_0$.
Let $L$ be the standard Liouvillean of the group $\tau_{H_0+\lambda V}$
and $\Phi\in\cP$ the standard representative of the KMS state $\mu_{H+\lambda V}$.
For $X,Y\in\mathfrak{X}$ one has
$$
( \rho_{H+X+\lambda V},Y)=\mu_{H+X+\lambda V}(Y\otimes\one)=
\frac{\langle \e^{-\beta(L+\pi(X))/2}\Phi,\pi(Y\otimes\one)\e^{-\beta(L+\pi(X))/2}\Phi\rangle}%
{\|\e^{-\beta(L+\pi(X))/2}\Phi\|^2}.
$$
Using Eq.\;\eqref{PhiPrime}, an explicit calculation yields that the derivative 
$F'(\lambda, H_0)$ of the function $F$ with respect to its second argument  is the 
symmetric linear map on $\mathfrak{X}$ given by
$$
( F'(\lambda,H_0)X,Y)
=-2\int_0^{\beta/2}\Re\langle \e^{-sL/2}\pi(\hat X\otimes\one)\Psi,
\e^{-sL/2}\pi(\hat Y\otimes\one)\Psi\rangle \d s,
$$
where $\hat X=X-\rho_{H_0+\lambda V}(X)\one$. Since $\hat X=0$ iff $X=0$, it 
follows that
$$
(F'(\lambda,H_0)X,X)
=-2\int_0^{\beta/2}\|\e^{-sL/2}\pi(\hat X\otimes\one)\Psi\|^2\d s<0,
$$
for all $0\not=X\in\mathfrak{X}$, and the implicit function theorem yields the conclusions of Proposition \ref{HTarget}.
\hfill\qed

\paragraph{Proof of Proposition \ref{roof}}  Suppose that $\sigma(\gamma)=0$. 
The weak-$\ast$ lower semicontinuity of relative entropy yields
\[
0=\sigma(\gamma)
=\lim_{t\to\infty}S(\omega_\i\circ\tau_{K_\gamma}^t|\rho_{K_\gamma}(t)\otimes\nu_\i)
\geq S(\mu_{K_\gamma}|\rho_\f\otimes\nu_\i),
\]
which implies $\mu_{K_\gamma}=\rho_\f\otimes\nu_\i$ and hence $\mu_{K_\gamma}\circ\tau_\cR^t=\mu_{K_\gamma}=\mu_{K_\gamma}\circ\tau_{K_\gamma}^t$
for all $t\in\rr$. It follows that
$$
\mu_{K\gamma}(\delta_\cR(A))=0=\mu_{K_\gamma}(\delta_\cR(A)+\i[K_\gamma,A])
$$
for all $A\in\Dom(\delta_\cR)$, from which we conclude that $K_\gamma$ belongs
to the centralizer of $\mu_{K_\gamma}$. It follows from the KMS property of $\mu_{K_\gamma}$ that $\tau_{K_\gamma}^t(K_\gamma)=K_\gamma$ for all $t\in\rr$
(see, e.g., Proposition~5.3.28 in~\cite{BR2}). 

For $\zeta\in\rr$, set
$S_\zeta=\e^{\zeta K_\gamma}/\mu_{K_\gamma}(\e^{2\zeta K_\gamma})^{1/2}$ 
and note that $\xi_\zeta(A)=\mu_{K_\gamma}(S_\zeta A S_\zeta)$ defines a 
state in $\mathcal N_{\omega_\i}$. The mixing property and the fact that
$\tau_{K_\gamma}^t(S_\zeta)=S_\zeta$ yield
$$
\mu_{K_\gamma}(A)=\lim_{t\to\infty}\xi_\zeta\circ\tau_{K_\gamma}^t(A)=
\lim_{t\to\infty}\mu_{K_\gamma}(S_\zeta\tau_{K_\gamma}^t(A)S_\zeta)=
\xi_\zeta(A),
$$
from which we conclude that $\mu_{K_\gamma}(S_\zeta A S_\zeta-A)=0$ for
all $A\in\cO$. Setting $A=S_\zeta^2-\one$ further yields
$\mu_{K_\gamma}((S_\zeta^2-\one)^2)=0$. Since $\mu_{K_\gamma}$ is faithful
we conclude that $S_\zeta^2=\one$ and hence that $K_\gamma$ is a multiple
of $\one$. This implies that $\delta_{K_\gamma}=\delta_\cR$ and contradicts
Assumption~{\bf A}. \qed


\paragraph{Proof of Theorem \ref{adiabatic}}
Denote by $L$ the standard Liouvillean of the group $\tau_{K(0)}$. Let $\Psi(0)$ be 
the standard vector 
representative of the KMS state $\mu_{K(0)}$. 
For $t\in[0,T]$, set
\[
L_T(t)=L+\pi(\widehat K_T(t))-J\pi(\widehat K_T(t))J,
\]
with $\widehat K_T(t)=K_T(t)-K_T(0)$. The family $\{W_T(t)\}_{t\in[0,T]}$ of 
unitary operators on $\fH$ satisfying 
\[
\i\partial_t W_T(t)=L_T(t)W_T(t), \qquad W_T(0)=I,
\]
implements the dynamics $\alpha_{K_T}$ and preserves the natural cone, i.e., 
\beq\label{Conalpha}
\pi(\alpha_{K_T}^t(A))=W_T^\ast(t)\pi(A)W_T(t),\qquad 
W_T(t)\cP\subset\cP,
\eeq
for all $t\in[0,T]$ and all $A\in\cO$.
With $\widehat{K}(\gamma)=K(\gamma)-K(0)$, the standard Liouvillean of the 
instantaneous dynamics $\tau_{K(\gamma)}$ is 
$$
L(\gamma)=L+\pi(\widehat K(\gamma))- J\pi(\widehat K(\gamma))J,
$$
and the standard representative of the KMS state $\mu_{K(\gamma)}$ is
\[
\Psi({\gamma})=\frac{\e^{-\beta(L+\pi(\widehat K(\gamma)))/2}\Psi(0)}%
{\|\e^{-\beta(L+\pi(\widehat K(\gamma)))/2}\Psi(0)\|}.
\]
By construction, the orthogonal projection
$$
P(\gamma)=|\Psi(\gamma)\rangle\langle\Psi(\gamma)|
$$
is such that $\Ran(P(\gamma))\subset\Ker(L(\gamma))$ for $\gamma\in[0,1]$. 
Moreover, Assumption {\bf B} implies $\Ran(P(\gamma))=\Ker(L(\gamma))$ for 
$\gamma\in]0,1[$. Since the function $]0,1[\ni \gamma\mapsto\widehat{K}(\gamma)$ is
$C^2$ in norm  with uniformly bounded first and second derivative, the expansion~\eqref{araki-exp}, the estimate~\eqref{araki-est}, and an 
obvious telescoping argument show that the map
\[
]0,1[\ni\gamma\mapsto P(\gamma)\in\cB(\fH)
\]
is also $C^2$ in norm with uniformly bounded first and second derivative.

One easily checks that the adiabatic evolution $\cW_T(t)$ defined by 
\[
\i\partial_t\cW_{T}(t)
=(L_T(t)+T^{-1}\i[\dot P(t/T),P(t/T)])\cW_{T}(t), 
\qquad \cW_{T}(0)=\one.
\]
intertwines $P(0)$ and $P(t/T)$, i.e., that
\beq\label{Intertw}
\cW_{T}(t)P(0)=P(t/T)\cW_{T}(t),
\eeq
holds for $t\in[0,T]$.

With these preliminaries, the Avron-Elgart adiabatic  theorem \cite{AE,Teu,ASF1} gives: 

\bet\label{adiabatic-teufel}
Suppose that Assumption {\bf B} holds. Then 
\[
\lim_{T\rightarrow \infty}\sup_{t\in[0,T]}\|W_T(t)-\cW_{T}(t)\|=0.
\]
\eet
For $\gamma\in[0,1]$, it follows from Eq.\;\eqref{Conalpha} that
\[
\mu_{K(0)}\circ\alpha_{K_T}^{\gamma T}(A)
=\langle W_T(\gamma T)\Omega,\pi(A)W_T(\gamma T)\Omega\rangle,
\]
while the intertwining relation~\eqref{Intertw} yields
\[
\mu_{K(\gamma)}(A)
=\langle \cW_{T}(\gamma T)\Omega,\pi(A)\cW_{T}(\gamma T)\Omega\rangle.
\]
Thus, we have the estimate
\[
|\mu_{K(0)}\circ\alpha_{K_T}^{\gamma T}(A)-\mu_{K(\gamma)}(A)|
\leq 2\|\A\|\sup_{t\in[0,T]}\|W_T(t)-\cW_{T}(t)\|,
\] 
which, together with Theorem~\ref{adiabatic-teufel}, yields 
Theorem \ref{adiabatic}.\hfill\qed
\bigskip

\paragraph{Proof of Proposition \ref{thm-td-1}}
\label{proof-thm-td-1}

We use the same notation as in the proof of Theorem \ref{adiabatic}.  Set $\Phi(\gamma)=\e^{-\beta(L+\pi(\widehat{K}(\gamma)))/2}\Psi(0)$. Araki's
perturbation formula yields
$$
S(\omega|\omega_{K(\gamma)})
=S(\omega|\omega_\i)+\beta\omega(\widehat{K}(\gamma))+\log\|\Phi(\gamma)\|^2,
$$
for any $\omega\in\cN_{\omega_\i}$. Setting $\omega=\omega_{\i/\f}$ and
$\gamma=0/1$ we derive $\|\Phi(0)\|=\|\Phi(1)\|=1$. Next, we claim that
\beq
\mu_{K(\gamma)}(\partial_\gamma K(\gamma))
=-\frac{1}{\beta}
\partial_\gamma\log\|\Phi(\gamma)\|^2,
\label{really}
\eeq
which clearly implies Proposition~\ref{thm-td-1}.

The identity
\[
\partial_\gamma\log\|\Phi(\gamma)\|^2
=\frac{\langle \partial_\gamma\Phi(\gamma),\Phi(\gamma)\rangle
+\langle \Phi(\gamma),\partial_\gamma\Phi(\gamma)\rangle}{\|\Phi(\gamma)\|^2},
\]	
implies that (\ref{really}) follows from
$$
\langle\Phi(\gamma),\partial_\gamma\Phi(\gamma)\rangle
=-\frac{\beta}{2}
\langle \Phi(\gamma),\pi(\partial_\gamma K(\gamma))\Phi(\gamma)\rangle.
$$
The last identity is a  direct consequence of Eq.\;\eqref{PhiPrime} and the fact
that $L(\gamma)\Phi(\gamma)=0$.\hfill\qed


\end{document}